\documentclass[sigconf]{acmart}
\pdfoutput=1

\setcopyright{none} 

\def\withComments{1} 
\usepackage{hyperref}
\usepackage{amsmath,amsthm,amsfonts,amssymb}
\usepackage{aliascnt}
\usepackage{courier}
\usepackage{ifpdf}
\usepackage[normalem]{ulem}
\usepackage{color}
\usepackage{subfigure}
\usepackage{multirow}
\usepackage[ruled,vlined,linesnumbered]{algorithm2e}
\usepackage{tabulary}
\newcolumntype{K}[1]{>{\centering\arraybackslash}p{#1}}
\newcolumntype{L}[1]{>{\raggedright\arraybackslash}p{#1}}
\usepackage{graphicx}

\ifdefined\withComments
\newcommand{\authnote}[2]{{\bf [{\color{red} #1's Note:} {\color{blue} #2}]}}
\else
\newcommand{\authnote}[2]{}
\fi

\newcommand{\eliad}[1]{\authnote{Eliad}{#1}}

\newcommand{\remove}[1]{}

\newcommand{\mparagraph}[1]{\ \\ \noindent{\bf{#1}}}

\def\AE{{AlterEve }}
\def\EConf{{EConfidante }}
\def\EC{{EConfidante }}
\newcommand{\ek}{\ensuremath{\mathsf{econf\_key}}}

\newcommand{\sdotfill}{\textcolor[rgb]{0.8,0.8,0.8}{\dotfill}} 

\newenvironment{algorithmm}{\begin{algo}}{\vspace{-\topsep}\end{algo}}

\newcommand{\ie}  {i.e.,\ }
\newcommand{\eg}  {e.g.,\ }

\newcommand{\MathAlg}[1]{\mathsf{#1}}

\remove{
\newaliascnt{lemma}{theorem}

\aliascntresetthe{lemma}
\crefname{lemma}{Lemma}{Lemmas}

\newaliascnt{claim}{theorem}

\aliascntresetthe{claim}
\crefname{claim}{Claim}{Claims}

\newaliascnt{corollary}{theorem}

\aliascntresetthe{corollary}
\crefname{corollary}{Corollary}{Corollaries}

\newaliascnt{construction}{theorem}

\aliascntresetthe{construction}
\crefname{construction}{Construction}{Constructions}

\newaliascnt{fact}{theorem}

\aliascntresetthe{fact}
\crefname{fact}{Fact}{Facts}

\newaliascnt{proposition}{theorem}

\aliascntresetthe{proposition}
\crefname{proposition}{Proposition}{Propositions}

\newaliascnt{conjecture}{theorem}

\aliascntresetthe{conjecture}
\crefname{conjecture}{Conjecture}{Conjectures}

\newaliascnt{definition}{theorem}

\aliascntresetthe{definition}
\crefname{definition}{Definition}{Definitions}

\newaliascnt{remark}{theorem}

\aliascntresetthe{remark}
\crefname{remark}{Remark}{Remarks}

\newaliascnt{notation}{theorem}

\aliascntresetthe{notation}
\crefname{notation}{Notation}{Notation}

\newaliascnt{proto}{theorem}

\newtheorem{proto}[proto]{Protocol}

\aliascntresetthe{proto}
\crefname{proto}{protocol}{protocols}
}

\newaliascnt{algo}{theorem}
\newtheorem{algo}[algo]{Algorithm}
\aliascntresetthe{algo}

\newaliascnt{expr}{theorem}
\newtheorem{expr}[expr]{Experiment}
\aliascntresetthe{expr}

\newaliascnt{property}{theorem}
\newtheorem{property}[property]{Property}
\aliascntresetthe{property}

\def\FullBox{$\Box$}
\def\qed{\ifmmode\qquad\FullBox\else{\unskip\nobreak\hfil
\penalty50\hskip1em\null\nobreak\hfil\FullBox
\parfillskip=0pt\finalhyphendemerits=0\endgraf}\fi}

\def\qedsketch{\ifmmode\Box\else{\unskip\nobreak\hfil
\penalty50\hskip1em\null\nobreak\hfil$\Box$
\parfillskip=0pt\finalhyphendemerits=0\endgraf}\fi}

\newcommand{\adminpk}{\mathsf{adm\_pk}}
\newcommand{\adminsk}{\mathsf{adm\_sk}}
\newcommand{\iaspk}{\mathsf{ias\_pk}}

\newcommand{\econfk}{\mathsf{econf\_key}}

\newcommand{\leases}{\mathsf{leases}}

\newcommand{\hash}{\mathsf{hash}}

\newcommand{\iasreport}{\mathsf{ias\_report}}

\newcommand{\mrenclave}{\mathsf{mr\_enclave}}

\newcommand{\expmrenclave}{\mathsf{expected\_mr\_enclave}}

\newcommand{\userdata}{\mathsf{user\_data}}
\newcommand{\agentid}{\mathsf{agent\_id}}
\newcommand{\adminid}{\mathsf{admin\_id}}

\newcommand{\econfenclaveso}{\mathsf{econf\_enclave.so}}

\newcommand{\econfmrenclave}{\mathsf{econf\_mr\_enclave}}

\makeatletter
\newcommand\xleftrightarrow[2][]{%
	\ext@arrow 9999{\longleftrightarrowfill@}{#1}{#2}}
\newcommand\longleftrightarrowfill@{%
	\arrowfill@\leftarrow\relbar\rightarrow}
\makeatother

\newcommand{\VerifyPublicInfo}{\MathAlg{VerPubInfo}}
\newcommand{\ComputeSharedKey}{\MathAlg{SharedKey}}

\newcommand{\pk}{\mathsf{pk}}
\newcommand{\sk}{\mathsf{sk}}

\newcommand{\agentpk}{\mathsf{agent\_pk}}
\newcommand{\agentsk}{\mathsf{agent\_sk}}

\newcommand{\clientdata}{\mathsf{data}}

\newcommand{\pubinfo}{\mathsf{pub\_info}}

\newcommand{\trustledger}{\mathcal{L}}

\begin{document}
\title{It Takes Two to \#MeToo - Using Enclaves to Build Autonomous Trusted Systems}

\author{Danny Harnik \hspace{0.07in} Paula Ta-Shma \hspace{0.07in} Eliad Tsfadia}
\affiliation{%
	\institution{IBM Research -- Haifa}
}
\email{ {dannyh, paula, eliadt}@il.ibm.com}

\begin{abstract}
We provide enhanced security against insider attacks in services that manage extremely sensitive data. One example is a \#MeToo use case where sexual harassment complaints are reported but only revealed when another complaint is filed against the same perpetrator. Such a service places tremendous trust
on service operators which our work aims to relieve.

To this end we introduce a new autonomous data management concept which transfers responsibility for the sensitive data from administrators to secure and verifiable hardware. The main idea is to manage all data access via a cluster of autonomous computation agents running inside Intel SGX enclaves.
These {\em EConfidante} agents share a secret data key which is unknown to any external entity, including the data service administrators, thus eliminating many opportunities for data exposure.
In this paper we describe a detailed design of the EConfidante system, its flow and how it is managed and implemented. Our \#MeToo design also uses an immutable distributed ledger which is built using components from a Blockchain framework. We implemented a proof of concept of our system for the \#MeToo use case and analyze its security properties and implementation details.
 \end{abstract}




\maketitle

\section{Introduction}\label{sec:intro}

The protection of sensitive data is a central security concern for any organization retaining and utilizing such data.
Indeed, there are numerous protection mechanisms, including encryption, access control and virtual or physical data isolation.
A key weakness is reliance on a centralized entity that is entrusted with the management of the IT system, its health, scalability and availability.
Typically, IT administrators have nearly limitless access to data or privileges to bypass or change the software for managing it. The existence of powerful privileged users in a system leaves an opening for foul play, whether by a rogue employee or via hacking and hijacking of secret privileged credentials.

History has shown that such trust can, and has been, breached, whether maliciously or not. Incidents of data theft by authorized persons have been abundant, with some high profile data breaches such as the Snowden leaks. Data leaks continue to plague various organizations, be it by a disgruntled employee, inadvertent error, or by hacking of accounts of individuals that have the credentials to override the security obstacles instated for data protection.\footnote{In fact, it has been documented that a majority of the security breaches have come from inside organizations.\cite{InsideAttacks15}} Finally, many of defence mechanisms are implemented in software and can be bypassed by administrators with direct access to the underlying hardware. This work attempts to significantly reduce the ability of rogue employees or hackers to extract sensitive data from systems that are charged with maintaining and handling sensitive information.

At the core of our work is a new {\em autonomous data management} concept -- transferring the responsibility for data from administrators to secure hardware. A good analogy would be entrusting data to a robot that is both autonomous yet always truthful to its cause. By Autonomous we mean that it cannot be influenced or inspected by any external party. A second property is that it does only what it was programmed to do, it is robotic in the non-creative sense of the word. 
 Our instantiation of the autonomous and trustworthy robot builds on new hardware paradigms for secure execution, and specifically on Intel's SGX technology \cite{SGX_Web}.

This approach conceptually has wide applicability as a security enhancement, but is especially beneficial for targeted applications that maintain highly sensitive data and in which, for the most part, the data never leaves the confines of the secure repository. One such example is a MeToo use case, in which sexual harassment complaints are registered in confidence under the understanding that they will not be reported to authorities unless there are at least two matching complaints against the same perpetrator. Such services already exist, e.g. Callisto~\cite{callisto} and ``One of One"~\cite{one-of-one}, but they put extremely high trust in their operators and administrators. This use case is discussed in depth in Section~\ref{sec:usecases} along with other relevant applications.

The basic idea is to have all sensitive data encrypted with a key that is known only to software running inside Intel SGX Enclaves. Enclaves are special computation modes in Intel Skylake processors that run all computations on encrypted data. Coupled with a remote attestation capability (a code and platform verification mechanism), the enclaves can serve as the desired robot for executing all data related operations. Our aim is to ensure that the key to the encrypted data only resides inside an {\em \EC agent} running in an enclave and therefore the encrypted data can only be useful if accessed via the enclave. In a sense, we are replacing the parts of the privileged software that actually handle the data by robotic code running in a secure execution environment.



\mparagraph{Contributions:} 
We present a detailed architecture and design of the \EC system and analyze the security properties it achieves.
We complement the cluster of \EC agents, which are responsible for keeping the encryption key secret, with additional mechanisms to ensure data survivability, immutability, auditing capabilities and control over the cluster.
We build an immutable distributed ledger in which the agents share data. The Ledger is built using HyperLedger Fabric, a framework for Blockchain applications, and is run by the system administrators that are entrusted with controlling the cluster. These administrators also run a lease based control ecosystem for creating, running and maintaining the agent cluster.
Our design is such that a single rogue administrator (or a configurable number of rogue administrators) cannot manipulate the data in the ledger or the makeup of the \EC cluster. We stress that the management's responsibility is to support and control the cluster, but even a fully malicious management cannot retrieve
the secret key from the \EC agents without breaking the security of the SGX technology.

We implemented a proof of concept of an \EC system for the MeToo use case built on Intel SGX hardware and using the HyperLedger Fabric code base. We report implementation details, lessons learned and performance attributes of our PoC.

\mparagraph{Paper Outline:}
Section~\ref{sec:usecases} discusses the MeToo use case as well as other interesting data services.
Section~\ref{sec:BG} presents background regarding the Intel SGX technology and HyperLedger Fabric.
In section~\ref{sec:design} we present a high level overview of our design and its goals while in Section~\ref{sec:details} we present a detailed account of this design.
The implementation is described in section~\ref{sec:implementation} and Section~\ref{sec:related} describes related work. Appendix~\ref{sec:sec} provides security properties of the MeToo implementation and Appendix~\ref{app:details} contains additional discussion on design and implementation.



\section{Use Cases}\label{sec:usecases}
In this paper we focus on {\bf protection of centrally managed sensitive data against the privileged user threat}, to provide additional protection against data leakage.
This section details several use cases. We focus on the \#MeToo use case, and also mention genomics and biometric database use cases.

\subsection{The \#MeToo Use Case}\label{sec:usecase-metoo}
Since October, the phrase ``Me Too'', originally coined by Tarana Burke and popularized as a hashtag by Alyssa Milano, has been tweeted over 1.7 million times. The hashtag went viral in the wake of sexual harassment allegations against Hollywood mogul Harvey Weinstein.
Why have victims been silent until now, and why has the \#MeToo campaign become so intensely viral?
Sexual assault victims risk significant damage and often obtain no clear benefit for providing their testimonies.
It takes two to \#MeToo and someone had to be first 
\#MeToo shows the immense power of bringing together victims of the same offender.

Recently software systems such as Callisto\cite{callisto} have been deployed to help victims of sexual assault record and possibly report offences. Callisto allows private documentation of the incident, with the option to release the records to authorities if the complaint matches another complaint for the same perpetrator.
Similarly, the `One of One' non-profit organization maintains a website and an encrypted repository of alleged sexual harassers and (only) uses this information to notify victims of the existence of matching complaints \cite{one-of-one}. Callisto uses algorithmic processes to perform the matching and notifications\cite{callisto-core} whereas for One of One this process is manual.
However in both cases, privileged administrators have access to complaints which requires considerable trust because this information could be hacked or leaked. \footnote{A new version of Callisto is under development which attempts to mitigate this\cite{rajan2018callisto} using a different approach than ours. We cover this in our Related Work section (Section~\ref{sec:related}).} Documented evidence of sexual harassment is extremely private data, and if leaked in an untimely fashion it can cause permanent lifelong damage. While efforts are made to secure the data, all systems today are susceptible to data leakage caused by human error and insider attacks, simply because there are humans involved with access to the data.

Beyond insider attacks, there is also a risk that data can be requested by court order using a subpoena process. The fact that Callisto users can document an assault but choose not to report it may have legal ramifications. Are the Callisto administrators considered to be in possession of this data? Could they be held responsible for refraining to take action to prevent further incidents? 
A unique feature of \EConf is that, unless there is a match, it is impossible to gain access to the data, even when all administrators are malicious.\footnote{Under our trust assumptions which are covered in Sections~\ref{sec:goals} and \ref{sec:sec}.}

\begin{figure}[h!]
\includegraphics[width=0.4\textwidth]{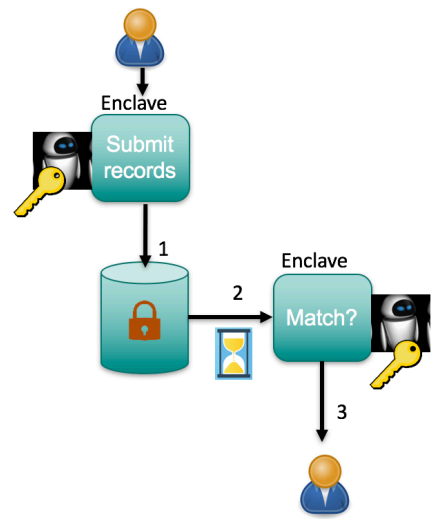}
\vspace{-.2in}
\caption{
Flows for the \#MeToo use case}\label{fig:metooflows}
\vspace{-.2in}
\end{figure}

Our main goal is to enhance the security of such systems so that unmatched allegations can be stored securely and indefinitely without ever being viewed by a human other than their author, while still guaranteeing them to be continuously matched against recent allegations.
In our work we do not address issues that concern registration and identity questions, and issues related to false accusations. We refer the reader to \cite{callisto, one-of-one} on such matters. We also do not define who the authorities are. Rather we consider the model of operation depicted in Figure ~\ref{fig:metooflows} and devise mechanisms to secure it. The main flows for this use case are:
\begin{enumerate}
\item User registers with \EConf and submits information about sexual assault, which is encrypted by \EConf using a unique key known only to \EConf and stored in the system.
\item Periodically an \EConf agent checks for matching perpetrators.
\item A periodic encrypted message is sent to the authorities containing any new matching complaints.
\end{enumerate}
A side note is that the message to the authorities is of constant length (even if empty), to avoid leaking information about whether recent matches were found.




\subsection{Biometric Databases}\label{sec:biometric}
In this use case, personal biometric data is stored by a central authority for means of identification of individuals.
Adding new biometric information to such a repository is a natural operation, for example, recording fingerprints of a new citizen. On the other hand, extracting biometric data should not be supported. Rather, the only information that should leave the database is an answer whether a given fingerprint matches an existing entry or not.

\subsection{The Genomics Use Case}\label{sec:usecase-genomics}
 An increasing number of B2B bioinformatics data platforms\cite{googlebio,amazonbio} and B2C services \cite{mog,23andme,myheritage} offer storage of genomic data and (platforms for) value added services such as genetic screening, ethnicity analysis and kindred search. While this data is extremely sensitive and its inadvertent leakage is irrevocable, platforms managing it have not been immune to hackers\cite{MHbreach}.
 Running these services where genomes are always encrypted and only accessed via \EC enclaves offers much stronger privacy to individuals adding their data to the service.
A key difference between this use case and MeToo include the sheer size of the data and amount of resources required to process it, and while memory and other limitations for Intel SGX are deterrents today, these barriers will likely be overcome over time.
Finally, a requirement of the genomics use case is to support an unanticipated set of genomic analytics capabilities. To handle such a use case the \EConf system should be extended to handle a flexible set of digitally signed and attested functionalities. This is similar to the functional encryption model with enclaves described in Iron\cite{Iron17}, where the role of \EConf would be to manage the encryption keys.

\section{Background}\label{sec:BG}
\subsection{SGX and Enclaves}\label{sec:SGX}
Intel Software Guard Extensions (SGX) \cite{MAB13SGX} is a set of instructions on some of Intel's latest X86 platforms which aims to allow application developers to run parts of their code in a protected, isolated fashion \cite{HMR13SGX}. These instructions create execution environments called {\em Enclaves} which run in secluded memory areas and use hardware based encryption to the content of memory used by the enclave process.
There are many details on SGX's implementation and usage (e.g. on the SGX web-site \cite{SGX_Web} or in detailed studies \cite{CD16}). Here we give a very brief overview while focusing on the central relevant properties.

At a high level, the two main features of SGX are Isolation and Attestation. Isolation is a process' ability to run in a secluded manner where other processes, even with high privileges, cannot read its memory pages or modify them. Attestation is the ability to prove to external parties a) that the code is running inside an Intel SGX enclave, and b) that specific code is being executed in the enclave. In addition, SGX supports a sealing operation which allows an enclave to store its encrypted state on persistent storage and later retrieve it and continue operation (on the same machine only).

\mparagraph{SGX Attestation:}
 The attestation process (described in \cite{AGJ13SGX}) entails the creation of an attestation {\em quote} of an enclave which is signed using a key that exists only in platforms with Intel SGX (the signature is an asymmetric signature scheme, and more precisely a ring signature called EPID~\cite{EPID10}). The quote signature can then be verified by contacting an Intel Attestation Service (IAS).
The attestation quote includes a {\em measurement} over the enclave to be verified. This measurement, denoted MRENCLAVE, is essentially a cryptographic hash of the enclave's memory pages (reflecting its code during its creation). The quote also includes various information about the platform, the creator of the enclave and most importantly a field for additional user data. This user data can be used to store a public key (generated by the enclave) for all future identification and communication with the enclave.
 Once an attestation process has completed, the enclave being attested has a publicly verifiable IAS report (which can be verified using an IAS public key). The IAS report includes the content of the quote and serves as proof that an enclave with the corresponding measurement is indeed executing in an Intel SGX platform and that it can be identified and communicated with using the corresponding enclave's public key.


\mparagraph{SGX usage limitations:}  There are strict limitations on the code running inside SGX enclaves, in particular, code needs to be written in C/C++ and system calls cannot be made. An enclave therefore is made out of code running inside the enclave, and wrapper code running outside the enclave, making {\em ECALLs} to functions within the enclave (defined in an "edl" file). This wrapper is usually referred to as the {\em untrusted} part of the SGX code. The untrusted part handles communication between the enclave and peripherals (e.g. network, storage).

The SGX framework also comes with significant limitations on the amount of memory an enclave is allocated, after which paging from the enclave designated memory area (the Enclave Page Cache) to the general memory needs to occur (hence affecting performance, see~\cite{Scone16,Eleos17, Hotcalls17}). Currently the limit of the enclave page cache is 128MB.

\mparagraph{Attacks on SGX:}
Although great efforts have been made to isolate an enclave process from its environment, numerous side channel attacks have been demonstrated,  (e.g.\cite{CD16,SWGMM17,BMDKCS17,GESM17cache,LS+17,WC+17,SGXSpectre18,SgxPectre18}). Mitigation of such attacks has been discussed at length in the aforementioned studies and we consider these beyond the scope of our work. Our use of enclaves adheres to best practice guidelines for avoiding side channel attacks and the main security claim assumes that code running in enclaves is indeed secure.

That said, our general approach to using SGX enclaves is that they should be an additional line of defence rather than replacing existing security mechanisms. By doing so we can assure that in case of an enclave security breach, the level of trust in the system falls back to the security mechanisms that predated the use of SGX enclaves. In our case this means that the security falls back to trust in the system administrators (as it was before introducing \EC).

\subsection{The HyperLedger Fabric Blockchain Framework}\label{sec:fabric}
HyperLedger Fabric (or simply Fabric) \cite{hyper, Hyperledger18} is a platform for distributed ledger solutions. It is one of the HyperLedger open source projects hosted by The Linux Foundation and is designed to support pluggable implementations of different components such as consensus and membership services. HyperLedger Fabric leverages container technology to host smart contracts called ``chaincode'' that comprise the application logic of the system.
We use the framework to maintain an immutable distributed ledger (which serves as shared storage) that can be queried for the latest state (thereby preventing replay attacks) and can be verified inside SGX enclaves.
See Section~\ref{sec:ledger} for details. 

\section{Design Overview}\label{sec:design}

\subsection{Security Goals}\label{sec:goals}

 As mentioned in the introduction,
our system targets mitigating insider attacks including attacks by one or more rogue administrators.
The specific security requirements vary from one use case to another, yet we put forward common mechanisms that have a wide applicability.

The central concept of our system is to rely on SGX technology to form a cluster of autonomous processes that manage data and all access to it. This is achieved by encrypting all sensitive data with a secret key (the \ek) which is known only inside the \EC enclaves, hence rendering data useless unless accessed via an enclave. We classify this as the {\em secrecy} property that our system provides and is the most important cornerstone of our system. However, there are additional security requirements that are needed and we discuss these as well.
The main security goals are listed below in order of importance.

\begin{itemize}
\item {\bf Secrecy:} The main goal is to provide secrecy for the data stored and handled by the system and our system provides strong guarantees based solely on the integrity of the Enclave hardware. Namely, the secrecy property holds even in face of an adversary that corrupts {\em all} of the system administrators.\footnote{Note that protection in face of corruption of {\em all} administrators assumes that the ability to upgrade enclave code requires additional permissions. See Section~\ref{sec:upgrade} for details.} Note that in case of an SGX security breach, the secrecy falls back to what it would be in a regular deployment. Namely, the data remains secure as long as no administrator is corrupted.

    The formal definition of data secrecy depends on the application since some applications are expected to divulge some data as part of their functionality. A general security statement that applies to all use cases therefore focuses on the secrecy of the cluster's key (\ek), or more precisely, states that the security of data encrypted with this key is not diminished even in the presence of malicious administrators.
    For the MeToo use case we guarantee the secrecy of all complaints. Complaints that have matches are sent to the authorities in an encrypted message, whereas the details of complaints that do not have matches are never decrypted outside an \EC agent. The formal guarantees for this use case will be discussed in Appendix~\ref{sec:sec}.

\item {\bf Immutability:} An important property for an \EC system is that data or state that was ingested by \EC agents cannot be changed or reversed. For the MeToo use-case this property is critical - it guarantees that complaints that were accepted by the system cannot be ignored, lost or erased. Our system is built to provide a guarantee that two matching complaints will indeed be reported to the authorities, as long as both received an ack from an \EC agent. This property is hard to achieve based on enclave technology, since enclaves have limited support for holding verifiable persistent state and may be rolled back to a previous state. Instead, we ensure this property by building an immutable {\em distributed ledger}, using the HyperLedger Fabric blockchain framework as the underlying technology. The distributed trust is guaranteed against any adversary that does not control at least $k$ out of $n$ administrators (for $k$ and $n$ of our choice).



\item {\bf Cluster control:} The key secrecy property is guaranteed by the \EC enclaves themselves, no matter how many of these are spawned. Still we invest efforts in controlling the number of running \EC agents that make up a cluster. There are several reasons for this: e.g., ensuring an orderly inception of the cluster, supporting orderly upgrades of enclave firmware/code, avoiding a split cluster or having {\em ``shadow agents"} -- legitimate \EC agents which are removed from the main cluster and used on the side by a rogue administrator to support enumeration attacks, side channel attacks or simply undetected access to the data service. Our design for this is based on a lease mechanism by which each \EC agent will only perform its operation if it receives a lease to do so by at least $k$ out of $n$ administrators.

    The lease mechanism also ties in with our auditing support in which the process of renewing a lease for an agent is restricted to agents who left a valid log and incident free log (and this log was not tampered with). See Section~\ref{sec:logging} for more information on the logging support.

\item {\bf Survivability:} Closely related to immutability, data survivability requires that both the data in the ledger is maintained, and the \EC agents that hold the $\econfk$ are available (otherwise the encrypted data becomes useless).  This property can be ensured even against a malicious administrator that attempts to launch a permanent denial of service attack, as long as $k$ out of $n$ enclaves behave properly (see details in Section~\ref{sec:survival}).
\end{itemize}

We emphasize that our design serves as a significant security enhancement to existing mechanisms but does not replace them altogether. For example, a MeToo application without enclaves will use strict access control policies to protect sensitive data and these policies and access control mechanisms must remain in place once the \EC cluster is introduced. This ensures that in case of a breach of the SGX technology (e.g. by new side channel attacks), the system falls back to existing mechanisms where administrators are entrusted with data secrecy.


\subsection{Architecture}\label{sec:arch}

We next describe the main building blocks comprising an \EConf solution as they are depicted in Figure~\ref{fig:arch}:
\begin{figure}[h!]
	\hspace{-0.2in}\includegraphics[trim=75 45 180 80,clip,width=0.5\textwidth]{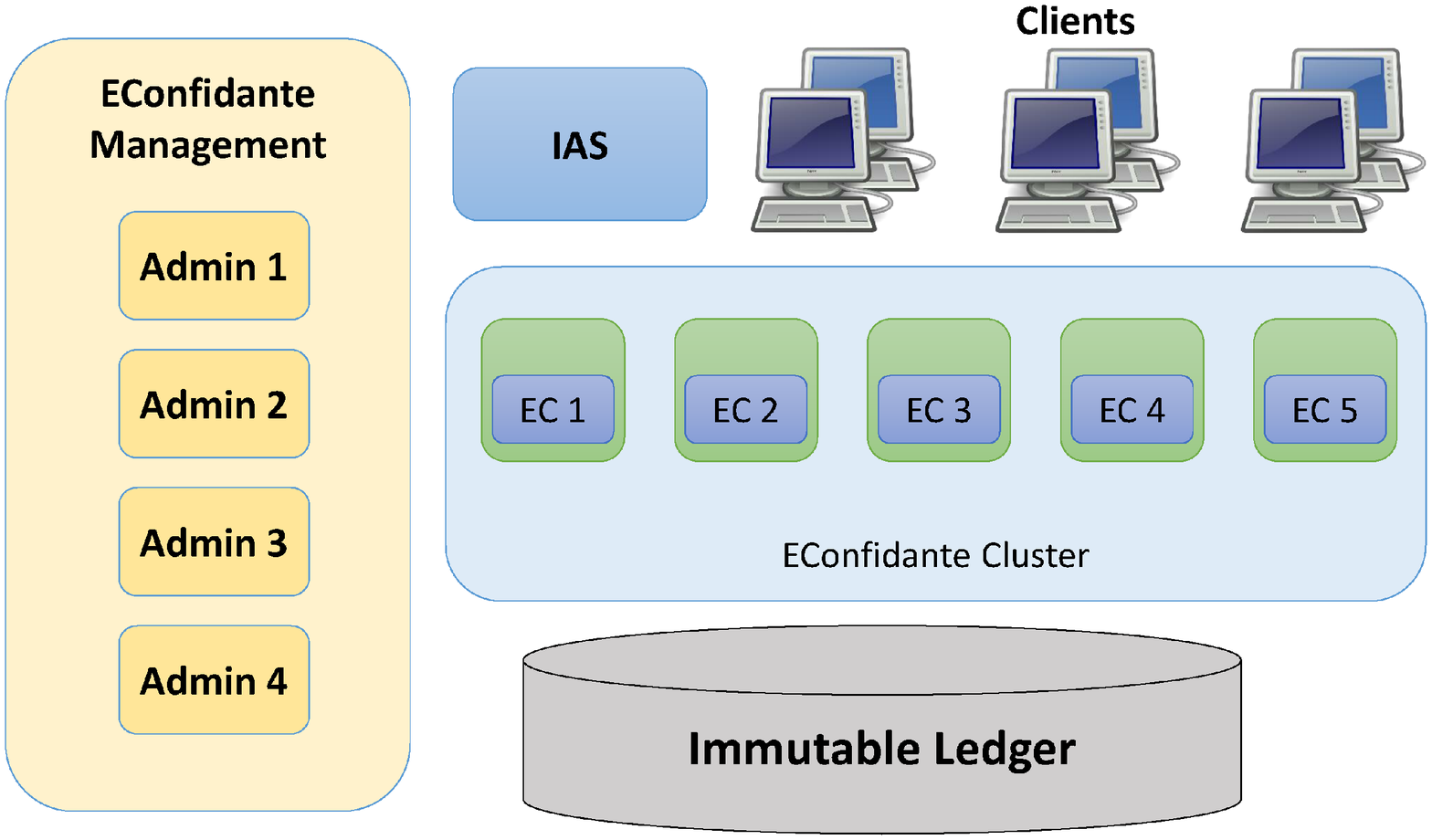}
	\vspace{-0.25in}
	\caption{
		The \EC system architecture.}\label{fig:arch}
\end{figure}

\mparagraph{\bf Administrators:}
At a high level, we view our system as being managed by a number of administrator entities. Each administrator has full access rights to manage the \EC cluster's resources (servers and network) so they can handle any maintenance and liveliness issues. But administrators should also have dedicated resources that cannot be accessed by the other administrators. These will be used for the immutable ledger and for producing leases. In particular, we assume that each administrator has a its own public and secret key-pair. In the following, we assume that $n$ is the number of administrators and we denote by $(\adminpk[i], \adminsk[i])$ the key-pair of the $i$'th administrator.

\remove{
although the key secrecy can still be maintained even w \eliad{what it means all powerful administrator? We should mention (here or somewhere else) that a malicious admin can access and tamper with any state in memory, on disk and over the network}. We also assume that there is a higher level of executives in charge of the service, and extremely sensitive operations would require their approval (such as during code upgrade of the \EConf agents) \eliad{Don't mention code upgrade. This is problematic. We need to mention somewhere that the ``key secrecy" property holds even when all administrators are malicious, but when enabling code-upgrade, this is not the case anymore}.
}

\mparagraph{\bf Intel Attestation service (IAS):}
This is the external service in which SGX enclaves are attested to and is provided by Intel~\cite{AGJ13SGX, EPID16SGX}.  
The attestation reports signed by the IAS can later be verified by all components in our system, and specifically by the \EC agents themselves. An IAS attestation report is constituted of an enclave quote and a signature on this quote signed by the IAS. The report can be verified using the IAS pubic key $\iaspk$.



\mparagraph{\bf EConfidante agents:}
As already mentioned, the backbone of \EConf is a collection of enclave agents sharing a (symmetric) secret key $\econfk$.  The fact that they run in enclaves ensures their security whether they run in a trusted or untrusted environment.\footnote{Although off premise enclaves may be more susceptible to side channel attacks.} There are two-types of agents: A leader agent and a regular agent. The leader agent is responsible for creating the $\econfk$ for the first time, and for sharing it with other (regular) agents. The other agents are responsible for receiving the $\econfk$ from some agent (leader or regular) and sharing the $\econfk$ with other agents which is critical for the survivability of the $\econfk$.  Moreover, they are responsible for the workload at hand. In the MeToo use-case, we divide these responsibilities into two new roles: A writer agent and a scanner agent. The writer agent is responsible for receiving clients' complaints and writing them into the immutable ledger. A scanner agent is responsible for periodically performing an offline scan of the ledger and finding new complaints that match, and for informing the authorities of such matches. An important property of our implementation is that all types of agents share the same code-base. An agent's role is only determined after it receives signed leases from $k$ out of $n$ administrators that specify whether it should act as a leader or as regular agent, and whether it should act as a writer or a scanner. Although the agents are (almost) stateless given a set of leases that specify their role, we use the sealing mechanism of SGX for saving an encrypted copy of $\econfk$ on disk. This enables the agent to recover after a reboot without receiving the $\econfk$ again from other agents.

\mparagraph{\bf The immutable ledger:}
This distributed ledger is managed and run by the administrators and is used as shared storage for \EC agents. It has the properties that an enclave reading from the ledger can be assured to get consistent up to date state, and in particular, the enclave will not accept old responses (that were previously accepted but no longer reflect the current state). We note that by design, the ledger is a distributed entity which requires interaction between several nodes for each update or query. As such, its performance is expected to be limited and this is indeed the case with our actual implementation (Section~\ref{sec:implementation}).
Therefore, the ledger is suitable for any information that requires immutability guarantees, but should be used sparsely due to its performance overhead. In the MeToo use case the ledger is used to store all complaints (this is reasonable since the number of complaints should be manageable). In other use cases (e.g. the biometric database) data can be stored in local storage or alternative shared storage.


\mparagraph{Clients:}
The clients represent users of the system. For example, in the MeToo  use-case they represent victims of incidents that report complaints.\footnote{In the MeToo use case clients should have some form of registration and identification. We view these logistics as out of the scope of our paper and refer the reader to the Callisto project~\cite{callisto,rajan2018callisto} for an example of handling such issues.}  Clients interact with the \EConf agents via secure communication based on the agent's public keys. There is no inherent limitation on how they are deployed, on what system and in what language they are implemented. In our implementation we provide clients with a means to verify an \EC agent's attestation report in order to ensure that they are indeed reporting their complaints to a genuine \EC agent running inside an SGX enclave.

\subsection{Putting it all Together}\label{sec:together}
We briefly describe how all these components fit together.
\EC agents receive complaints from clients and register these in the ledger. Periodically (e.g. once a day), a scanner agent retrieves all complaints from the ledger and scans for matches, and any new matches are written in an encrypted message which is sent to the authorities. The message is of fixed size, so it does not leak information in encrypted form. \EC agents are also responsible for handing the $\econfk$ to their fellow agents, ensuring the liveliness of the key.

\EC agents are programmed to run only if they have up to date leases from a sufficient number of administrators.
The administrators managing the \EC agents are responsible for controlling the cluster and ensuring its liveliness.
In regular operation, there is a fixed set of agents and the administrators have automatic scripts for lease renewal.
However in cases of cluster growth, or permanent node/enclave failure, a new enclave needs to be created.
This is done by a single administrator but requires the involvement of other administrators in creating leases for the new enclave ($k$ out of $n$ administrators are sufficient). The administrators also run the distributed ledger and are in charge of cluster inception.

The main components and flow of our design are described in detail in the next section.
Due to space limitations, the description of additional mechanisms for key survivability, software upgrades, scanner functionality and logging are described in Appendix~\ref{app:details}.

\section{Details and Flows}\label{sec:details}
\begin{table*}[ht!]
	\centering
	\begin{tabular}{|c|c|c|}
		\hline
		Parameter & Explanation & Initialization phase\\
		\hline
		
		$\iaspk$ & Intel Report Signing Public Key for Attestation & \multirow{3}{*}{ Hard coded} \\
		$\adminpk[1],\ldots,\adminpk[n]$ & Administrators' Public Keys & \\
		\hline
		$\econfmrenclave$ & Measurement of the agent-enclave's code & After Compilation \\
		\hline
		$\agentpk, \agentsk$ & Self public-secret key pair & \multirow{2}{*}{Attestation Phase} \\
		$\agentid$ & hash over $\agentpk$ & \\
		$\iasreport$ & Agent IAS report that bind its $\mrenclave$ and $\agentpk$. &   \\
		\hline
		$\ell_{id}[1],\ldots,\ell_{id}[n]$ & Leases from all admins for agent with $agent\_id = id$ & Setup Phase \\
		$\econfk$ & The \EConf Secret Key & \\
		\hline
	\end{tabular}
	\vspace{2mm}
	\caption{\EConf agent main parameters and the phases at which they are obtained.}\label{tab:params}
\end{table*}

In this section we dive into the details of our system and describe the various data flows between the various components. This includes the initial cluster inception, the attestation phase, key transfer protocol and the client agent interactions.

\mparagraph{Initial \EC Agent Code Creation:}
 As in most cases with enclaves, a base assumption is that the code for the \EC enclaves is developed, tested, audited, compiled and agreed on and we can assume that it is correct. Before compilation, some important keys are hard coded into the enclave code.
Then, the code is compiled once, and the resulting shared-object file (denote by $\econfenclaveso$) is published together with its $\mrenclave$ measurement (denote by $\econfmrenclave$).

In Table \ref{tab:params} we list the main security parameters held by the \EC agent together with the phase in which they are obtained.
In particular the IAS's public key $\iaspk$ and the administrator's public-keys $\adminpk[1],...,$ $\adminpk[n]$ are hard coded before compilation.

\begin{figure}[h]
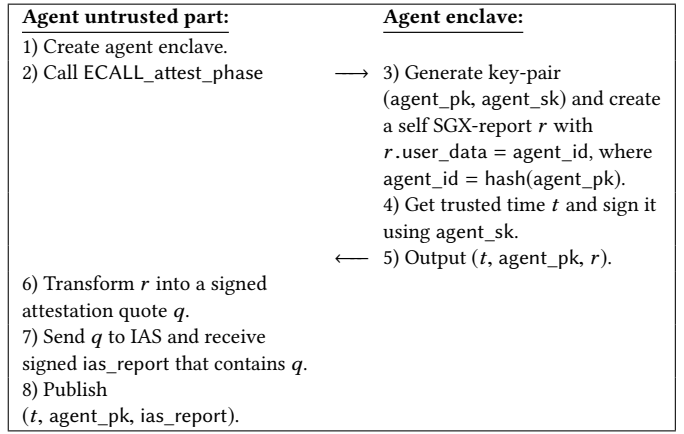

	\begin{center}
\begin{small}
		\begin{tabular}{|L{3.8cm}K{0.3cm}L{3.7cm}|}
			\hline
			{\bf \underline{Agent untrusted part:}} & \ &{\bf \underline{Agent enclave:}}  \\
			1) Create agent enclave.&  & \\
			2) Call $\mathsf{ECALL\_attest\_phase}$ & $\xrightarrow{\hspace*{0.3cm}}$ & 3) Generate key-pair $(\agentpk,\agentsk)$ and create a self SGX-report $r$ with $r.\userdata = \agentid$, where $\agentid = \hash(\agentpk)$.\\	
			& & 4) Get trusted time $t$ and sign it using $\agentsk$.\\
			& $\xleftarrow{\hspace*{0.3cm}}$ & 5) Output $(t, \agentpk, r)$.\\
			6) Transform $r$ into a signed attestation quote $q$. &&\\
			7) Send $q$ to IAS and receive signed $\iasreport$ that contains $q$. &&\\
			8) Publish $(t, \agentpk, \iasreport)$. &&\\
			\hline
		\end{tabular}
		\caption{\label{fig:attest_phase} The flow of agent creation and attestation.}
\end{small}
	\end{center}
\end{figure}

\mparagraph{Agent Creation and Attestation:}
We next describe the process by which an administrator creates a new \EC agent. The administrator first copies the agent binaries (including the file $\econfenclaveso$) to an SGX supported machine and then executes the agent. The operations that occur upon running the agent are described in Figure \ref{fig:attest_phase}.
Recall that an SGX application is made out of two parts: the enclave and the untrusted part. The untrusted part creates the agent-enclave and starts the attestation-phase. A key-pair $(\agentpk,\agentsk)$ is generated inside the enclave and the agent creates and publishes a publicly verifiable report (IAS report) that binds the value of $\agentpk$ into it by storing its $\agentid$ inside the user data of the signed attestation quote, where $\agentid = \hash(\agentpk)$. In addition, the agent also signs and publishes a trusted-time that has been generated inside the enclave using a special Intel function sgx\_get\_trusted\_time.\footnote{The trusted-time is a concatenation of the time itself (8 bytes) together with a time-nonce (32 bytes) provided by sgx\_get\_trusted\_time.} This trusted time will later be used for producing time-limited leases from the administrators.

\subsection{Operation Flows}\label{sec:flows}

\subsubsection{\bf Cluster Inception}\label{sec:cluster-inception}

\
The inception of an \EConf cluster is susceptible to various attacks such as the creation of a split cluster. Due to this vulnerability, we take a cautious approach for cluster inception by which only a fully successful setup is acceptable and any failure results in full abortion of the process and restart from scratch.

\begin{table*}[ht!]
	\centering
	\begin{tabular}{|c|c|c|c|c|c|c|c|}
		\hline
		$\agentid$ & $\adminid$ & $\mathsf{role}$ & $\mathsf{creation\_time}$ & $\mathsf{base\_trusted\_time}$ & $\mathsf{expiration}$ & $\mathsf{sig\_len}$ & $\mathsf{sig}$\\
		\hline
		32 bytes & 4 bytes & 4 bytes & 8 bytes & 40 bytes & 8 bytes & 4 bytes & $\mathsf{sig\_len}$ bytes\\
		\hline
	\end{tabular}
	\vspace{2mm}
	\caption{The structure of a signed lease}\label{tab:leases_struct}
\end{table*}

In this phase the secret key $\econfk$ should be generated inside a single \EC agent enclave and then shared with all other \EC agents. However, it needs to be decided which agent actually generates the key from a set of agents that must have the exact same enclave code (which is a crucial property for our solution as we argue next). Rather than implementing complex leader election code inside the agent enclaves, we elect to have this process handled at the administrator level using the lease mechanism.
After the attestation phase, the agent waits until it receives signed leases from at least $k$ out of the $n$ administrators. Each administrator $i \in \{1,\ldots,n\}$ should send a signed lease $\ell_{id}[i]$ using its secret key $\adminsk[i]$. As described in Table~\ref{tab:leases_struct}, a lease should contain the admin-id $i$, the agent-id $id$ (a hash over the agent's public key), the role of the agent (whether it should be a leader or not and whether it should be a writer or scanner), the agent's trusted time and an expiration time (\eg one day). In addition, a lease also contains its creation time (with respect to the administrator's clock) which can be used for additional non-enclave's verifications (\eg by the untrusted part of an agent in the key-transfer protocol, and by client software in the client-agent interaction).
When the agent-enclave receives enough leases with valid signatures (at least $k$), it is now allowed to operate as long as the expiration time of at least $k$ leases has not yet been reached (with respect to its trusted time). If the role that has been specified in all $k$ leases is a leader role, then this is the enclave that randomly generates the $\econfk$. Non-leader agents receive an IP address of an agent that holds the $\econfk$, and then the two agents start interacting in a key-transfer protocol, as described in Section \ref{sec:key-trans}. At the end of this phase all agents hold the $\econfk$ in their enclave's (encrypted) memory. At this point, the enclave stores its key-pair $(\agentpk,\agentsk)$ and $\econfk$ using the SGX sealing mechanism. This creates an encrypted file on disk that can be used by the agent to recover after a reboot. In addition, all the received leases are also stored on disk (without encryption, since they do not contain sensitive information). Once a lease has expired, the administrator should renew the lease. In Section~\ref{sec:logging} we suggest a way to decide whether to renew a lease based on our logging mechanism.

\subsubsection{\bf Key Transfer Between Agents}\label{sec:key-trans}

The most critical part in the life cycle of our system is the ability to securely transfer the $\econfk$ from one agent-enclave to another.
This is the only situation in which the $\econfk$ is transferred to another entity.
The receiver must only receive the $\econfk$ from a trusted sender (to prevent a fake $\econfk$). More importantly, the sender must only securely transmit its $\econfk$ to a trusted receiver. Figure~\ref{fig:transfer_econf_key_flow} describes the flow of our key-transfer protocol. At the first step, the agents send each other all their publicly verifiable information. This includes their $\agentpk$ and $\mrenclave$ measurement together with their $\iasreport$ that binds these values into it, and also includes signed $\leases$ from at least $k$ administrators. Next, each agent verifies the information of the other agent (using its own $\mrenclave$ as the $\expmrenclave$) and computes a (shared) secret-key based on its own secret-key and the other agent's public-key.\footnote{As mentioned in Section~\ref{sec:implementation}, since we use Elliptic-Curve's keys for the agent's key-pairs, such computation of shared secret key can be based on the Elliptic-Curve Diffie-Helman (ECDH) computation.} Finally, the sending agent uses the shared secret-key to send $\econfk$ to the receiving agent using authenticated encryption. Observe that the verification step must be executed inside the agent-enclave and not in the agent's untrusted part. Therefore, we implemented verification code that can be executed inside SGX enclaves. The following pseudo-code describes the verification steps:
\begin{algorithmm}[$\VerifyPublicInfo$]
	\item[Input:] $\agentpk$, $\iasreport$, $\leases$ and $\expmrenclave$.
	\item[Operation:]~
	\begin{enumerate}
		\item Let $\agentid = \hash(\agentpk)$.
		\item Verify that at least $k$ leases $\ell_{id}[i]$ have:
		\begin{enumerate}
			\item Correct signature (with respect to $\adminpk[i]$).
			\item $id = \agentid$.
		\end{enumerate}
		\item Verify the signature of $\iasreport$ using $\iaspk$.
		\item Extract the quote $q$ from $\iasreport$ and verify that:
		\begin{enumerate}
			\item $q.\userdata = \agentid$.
			\item $q.\mrenclave = \expmrenclave$.
		\end{enumerate}
	\end{enumerate}
\end{algorithmm}

\begin{figure}[h]
	\centering
	\begin{center}
	\begin{small}
		\begin{tabular}{|L{0.1cm}K{3.5cm}K{0.3cm}K{3.5cm}|}
			\hline
			& {\bf \underline{Sender Agent:}} & \ &{\bf \underline{Receiver Agent:}}  \\
			& Input: $(\pk_{s}, \sk_{s})$, $\iasreport_{s}$, $\leases_s$, $\econfk$ & &
			Input: $(\pk_{r}, \sk_{r})$, $\iasreport_{r}$, $\leases_r$ \\
			&&& \\
			1)& Send $\pubinfo_{s} :=$ $(\pk_{s},\iasreport_{s},\leases_s)$ & $\xleftrightarrow{\hspace*{0.3cm}}$ & Send $\pubinfo_{r} :=$ $(\pk_{r},\iasreport_{r},\leases_r)$\\
			&&&\\
			2)& Run $\VerifyPublicInfo(\pubinfo_{r}, $ && Run $\VerifyPublicInfo(\pubinfo_{s}, $\\
			&$\mathsf{self\_mr\_enclave})$ && $\mathsf{self\_mr\_enclave})$\\
			&&&\\
			3)& $k := \ComputeSharedKey(\sk_{s}, \pk_{r})$ & & $k := \ComputeSharedKey(\sk_{r}, \pk_{s})$\\
			&&&\\
			4)& Send $\mathsf{enc\_key}$, where $\mathsf{enc\_key} :=$ $Enc_{k}(\econfk)$.
			& $\xrightarrow{\hspace*{0.3cm}}$ & 5) $\econfk :=$ $ Dec_{k}(\mathsf{enc\_key})$\\
			\hline
		\end{tabular}
		\caption{\label{fig:transfer_econf_key_flow} Transfer of $\econfk$ between Cluster Agents}
	\end{small}
	\end{center}
\end{figure}

\remove{
\begin{figure}[h]
	\centering
	\begin{center}
		\begin{tabular}{|K{3.9cm}K{0.3cm}L{3.9cm}|}
			\hline
			{\bf \underline{Sender Agent:}} & \ &{\bf \underline{Receiver Agent:}}  \\
			Input: $(\pk_{s}, \sk_{s})$, $\iasreport_{s}$, $\leases_s$, $\econfk$ & &
			Input: $(\pk_{r}, \sk_{r})$, $\iasreport_{r}$, $\leases_r$ \\
			&& \\
			1) Send $\pubinfo_{s} :=$ $(\pk_{s},\iasreport_{s},\leases_s)$ & $\xrightarrow{\hspace*{0.3cm}}$ & \\	
			& & 2) Run $\VerifyPublicInfo(\pubinfo_{s}, $\\
			& & $\mathsf{self\_mr\_enclave})$\\
			& $\xleftarrow{\hspace*{0.3cm}}$ &
			3) Send $\pubinfo_{r} :=$ $(\pk_{r},\iasreport_{r},\leases_r)$\\
			4) Run $\VerifyPublicInfo(\pubinfo_{r}, $\\
			$\mathsf{self\_mr\_enclave})$\\
			5) $k := \ComputeSharedKey(\sk_{s}, \pk_{r})$\\
			6) Send $\mathsf{enc\_key}$, where $\mathsf{enc\_key} :=$ $Enc_{k}(\econfk)$.
			& $\xrightarrow{\hspace*{0.3cm}}$ \\
			&& 7) $k := \ComputeSharedKey(\sk_{r}, \pk_{s})$\\
			&& 8) Set $\econfk := Dec_{k}(\mathsf{enc\_key})$\\
			\hline
		\end{tabular}
		\caption{\label{fig:transfer_econf_key_flow_old} Transfer of $\econfk$ between Cluster Agents}
	\end{center}
\end{figure}
}

\subsubsection{\bf Client - \EC Agent Interactions}\label{sec:client_complaint}
\

\begin{figure}[h]
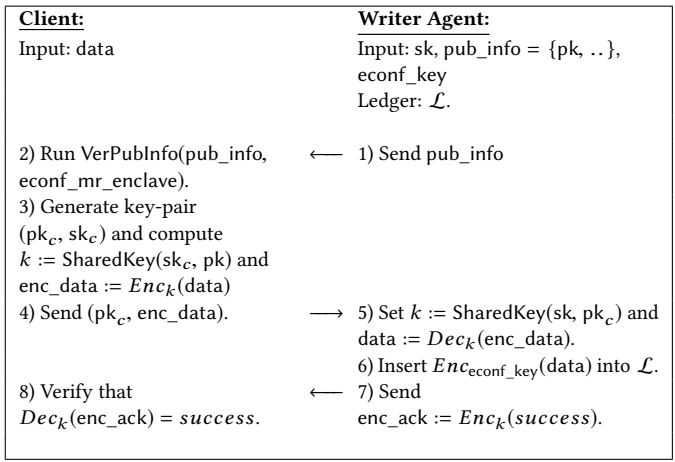

	\begin{center}
	\begin{small}
		\begin{tabular}{|L{3.5cm}K{0.3cm}L{4.0cm}|}
			\hline
			{\bf \underline{Client:}} & \ &{\bf \underline{Writer Agent:}}  \\
			Input: $\clientdata$ & &
			Input: $\sk$, $\pubinfo = \{\pk,..\}$, $\econfk$\\
			&& Ledger: $\trustledger$.\\
			&& \\
			2) Run $\VerifyPublicInfo(\pubinfo,$ $\econfmrenclave)$. & $\xleftarrow{\hspace*{0.3cm}}$ & 1) Send $\pubinfo$ \\
			3) Generate key-pair $(\pk_{c},\sk_{c})$ and compute $k := \ComputeSharedKey(\sk_{c}, \pk)$ and $\mathsf{enc\_data} := Enc_{k}(\clientdata)$ & & \\
			4) Send $(\pk_{c},\mathsf{enc\_data})$.& $\xrightarrow{\hspace*{0.3cm}}$ & 5) Set $k := \ComputeSharedKey(\sk, \pk_{c})$ and $\clientdata := Dec_{k}(\mathsf{enc\_data})$.\\
			& & 6) Insert $Enc_{\econfk}(\clientdata)$ into $\trustledger$.\\
			8) Verify that $Dec_{k}(\mathsf{enc\_ack}) = success$.& $\xleftarrow{\hspace*{0.3cm}}$ & 7) Send $\mathsf{enc\_ack} := Enc_{k}(success)$.\\
			& & \\
			\hline
		\end{tabular}
		\caption{\label{fig:client_to_agent_flow} Receiving Sensitive Data from a Client.}
	\end{small}
	\end{center}
\end{figure}

A client is a user of the \EC system that wants to interact with an \EC agent. In the MeToo use-case, this interaction is the reporting of a complaint. We describe the flow of a client that sends sensitive data to an \EC agent, which in turn stores it in the Ledger. The client connects to one of the \EConf writer agents, and requests the agent's $\agentpk$, its $\iasreport$ and leases. After the client verifies these, it generates an authenticated secure channel with the enclave and sends the data through it. The \EC agent re-encrypts the data using the $\econfk$ and writes it to the Ledger. Finally, after the agent verifies that the data was registered in the Ledger, it sends an acknowledgment to the client on a successful write. Figure~\ref{fig:client_to_agent_flow} gives a detailed description of the protocol flow.

\subsection{The Immutable Ledger}\label{sec:ledger}

We have two related reasons for using an immutable ledger. The first is our necessity to maintain a distributed database that is consistent across different users view and in particular cannot be changed or rolled back to a previous state. The second is that Intel SGX cannot prevent rollback of its storage state and is generally susceptible to replay attacks.\footnote{There is a limited support preventing rollback of SGX storage based on keeping a monotonic counter in an on-board non-volatile memory \cite{SGX_MC}. However this currently has some serious limitations such as hard limits on the number of updates to this counter, bad performance and limited support across different OS and platforms.}

There are several know approaches to handle these issues, and solving one issue will also help solving the other.
For example, there are several works that address the rollback issue by distributed trust. For example, ROTE \cite{ROTE17} distributes the state between a number of enclaves, and Brandenburger et al.\ \cite{BCLK17} relies on the clients of the system to jointly maintain a small state. We also choose to distribute state information and separate this responsibility from the \EC enclaves. This choice is made since the enclave technology is not essential to keeping a distributed state and having the joint state handled outside the enclaves (and just verified in the enclaves) relieves the enclaves from much of the complexity in maintaining the ledger and keep their TCB as small as possible. In our solution we create an administrator managed distributed ledger ensuring the following key property: a state written and acknowledged by the ledger is never erased. This in turn means that any enclave (or any other entity) reading from the ledger will never accept an old state (even if it was accepted before).

\mparagraph{Using HyperLedger Fabric:} Rather than building our own dedicated solution for this, we use an existing blockchain infrastructure, specifically, the HyperLedger Fabric project \cite{hyper,Hyperledger18} which is an open source project targeted for enterprise blockchain applications. We note that in a sense, the Fabric project is an overshoot for what we are doing -- it provides additional functionality that we do not require such as transaction ordering and validation. However, it presents an appealing development strategy for achieving the goals that we need, mainly, immutability and security and it natively handles membership services and networking.

We next describe our immutable ledger implementation based on HyperLedger Fabric and describe the guarantees that it provides.
Fabric presents a hierarchical structure of {\em organizations}, each of which maintains nodes consisting of {\em peers} \ and/or {\em clients}. The peers are responsible for maintaining the ledger state, while clients are hubs for handling requests (UPDATE and GET requests in our scenario).
Peers are programmed using so called {\em chaincode} which is the logic of what forms a legal transaction and what operations each transaction triggers. Bear in mind that Fabric is targeting Blockchain applications so a typical chaincode deals with monetary transactions, but the transactions for our use case are very simple and basically amount to UPDATE which adds a new entry to the ledger and GET which returns the current ledger state (this could be a specific entry or returning the entire ledger).
In addition, a distributed ledger has an associated {\em endorsement policy} which defines what is the set of peers that is sufficient for achieving a consensus. Examples of endorsement policies could be asking for at least a majority of the peers, or at least one peer from any organization (endorsement policies could vary for update vs.\ read operations).

\mparagraph{The trust model:}
It should be noted that our trust model differs fundamentally from what is typical in a standard fabric deployment.
In a typical deployment, each organization places full trust in the peers and clients of that specific organization and builds its security guarantees based on this local trust. In our case, on the other hand, only the \EC enclaves are fully trusted. The other entities in the system are susceptible to corruption, as is the network connecting between these entities and the enclaves. The trust assumption that we put is that at least $k$ out of $n$ administrators are not corrupted, and therefore receiving an endorsement from $k$ entities belonging to different administrators is enough to establish trust.
Technically speaking, we define each {\em administrator} in the \EC system as an {\em organization} in the HyperLedger Fabric setting. This means that each administrator holds a root certificate and from it can sign on certificates of peers and clients running under its domain (each administrator can run any number of peers or clients). The endorsement policy we require is to have at least $1$ peer from $n-k+1$ distinct organizations (administrators).

\begin{figure}[h!]
	\hspace{-0.2in}\includegraphics[clip,width=0.5\textwidth]{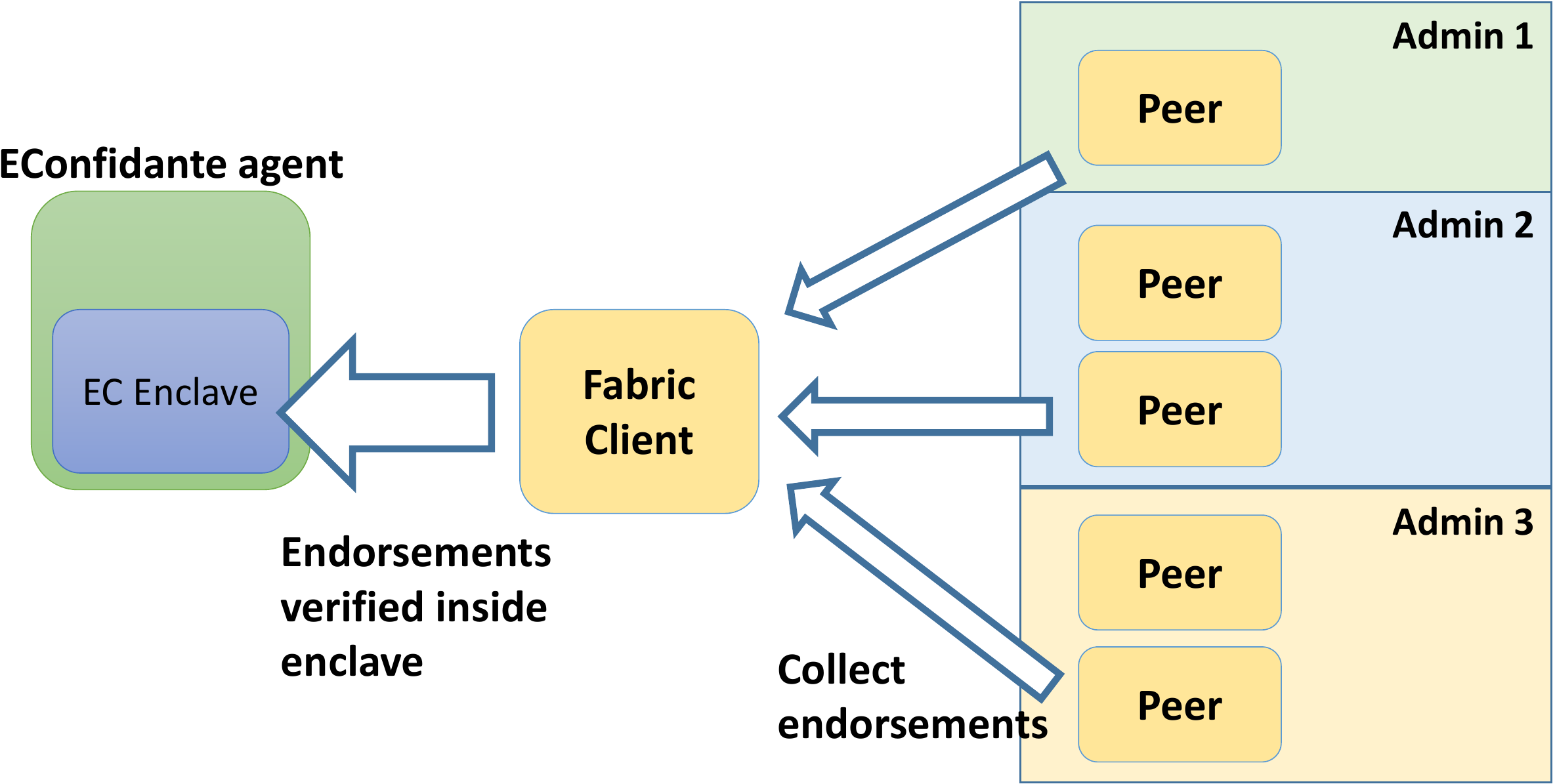}
	\vspace{-0.1in}
	\caption{
		A typical get operation flow in the Ledger}\label{fig:ledger}
\end{figure}

\mparagraph{The ledger operation:}
In practice, the Fabric flow is that each UPDATE to the ledger arrives at a fabric client which circulates this update among the peers and collects a sufficient number of {\em endorsements} to satisfy the endorsement policy. Each endorsement is basically a signature of a peer on the content of the update. Once enough endorsements are collected, they are packaged and sent to an {\em ordering service} which verifies the endorsements, sets an order to updates, signs off on it and resends the update to the peers to be incorporated to their local storage.
An ensuing GET follows a similar flow in which a client requests endorsements from the peers on their most recent state.
This is depicted in Figure~\ref{fig:ledger}. Once a sufficient amount of peers send endorsements to the same state, then this state is returned.

In our trust model, the \EC enclaves are the only fully trusted entities. All clients and peers could be potentially corrupted and therefore any response from the clients or peers need to be verified inside the \EC agent enclaves.
Basically this means that UPDATEs are issued by enclaves but performed by the agent helper code and a Fabric client. However, this UPDATE command is only acknowledge by the \EC agent by issuing a subsequent GET command that verifies that this UPDATE was indeed committed to the ledger. The result of the GET command along with all the endorsements from the peers are delivered back to the enclave and endorsement's validity is verified inside the enclave.
In addition, each GET request is accompanied with a nonce that is generated inside the enclave. This nonce is an integrated as part of the peer endorsements and allows the enclave to verify the liveliness of the response (this is key to avoiding the aforementioned rollback attacks).

\section{Implementation}\label{sec:implementation}

\remove{
\eliad{lines of code of interesting parts}

We measured total physical source lines of code using sloccount.
Total lines of our agent-enclave: 1229. This consists of:
attestation-phase: 163.
key-transfer code: 263.
processing admin leases code: 177.
processing client's data: 142.
sealing code: 196.

\eliad{sizes of IAS\_report and leases and blockchain entries?}

quote size: 436 bytes, signature:  680 bytes

IAS report total size: 5267 bytes.
IAS report consists of body, signature (base64) and cert chain.
body size:  1157 bytes, signature size: 344 bytes, cert chain size: 3766 bytes.

agent public key size: 158 bytes (EC-521).

lease total size: 160 bytes.
lease content: 88 bytes. signature size: 72 bytes (EC-256).

\eliad{timing measurements of interesting flows}

Attestation-phase time (including communication with IAS): $\approx 3$ sec.
Key-transfer protocol (without networking overhead): $\approx 50$ ms.
Client to agent flow (without networking and blockchain overhead): $\approx 25$ ms.

\eliad{Discussions: Attestation - mention that Intel's code is not good enough and we implemented an alternative that allowed enclave-to-enclave attestation, and tell that we have open source (anonymous)}

\eliad{Discussion 2: the problem with trusted time}
}

We implemented a proof of concept of the \EConf system including agents, administrators, clients and ledger. Our implementation was developed in a Linux Ubuntu release 16.04 environment using the Intel SGX SDK for Linux version 2.0 and tested on a Lenovo P50-series laptops with an Intel Core i7-6820HQ CPU @ 2.70GHz and 32GB memory.

For the cryptographic operations we use SGXSSL, available at the intel-sgx-ssl git repository\cite{sgx-ssl}. For the hash function we use $SHA256$ and for the symmetric encryption operations, we use AES256-GCM as an encrypt-then-mac method~\cite{BN00} to ensure authentication and IND-CCA2 security of the encryption scheme. We implemented the agent's key-pair $(\agentpk,\agentsk)$ using Elliptic-Curve 521-bit (EC521) keys.\footnote{Although there are no known attack on EC256, we chose to use 521-bit for the agents' keys in order to emphasize their sensitivity.} We use elliptic-curve Diffie-Helman (ECDH) for computing the shared-key as described in Figure~\ref{fig:transfer_econf_key_flow} and \ref{fig:client_to_agent_flow}, which results in a shared 256-bit symmetric key. In addition, we implemented the administrator's key-pairs $(\adminpk[i],\adminsk[i])$ using EC256 keys and for the administrator's signatures we use the elliptic-curve digital signature algorithm (ECDSA) with 256 bits.

Our overall enclave's code consists of $\approx 1700$ physical source lines of code (calculated by sloccount), where the breakdown of the lines of code of the  main modules is described in Table~\ref{tab:lines_of_code}.

\begin{table}[ht!]
	\centering
	\begin{tabular}{|c|c|}
		\hline
		Module & Lines of Code\\
		\hline
		Verifying cert\_chain and signature & 120\\
		\hline
		Verifying IAS-report & 187\\
		\hline
		Initialization and Attestation & 163 \\
		\hline
		Key-Transfer Protocol & 263 \\
		\hline
		Lease Processing & 177\\
		\hline
		Client's Complaint Processing & 180\\
		\hline
		Scanner functionality & 245\\
		\hline
		Sealing & 196\\
		\hline
		Logging & 80\\
		\hline
	\end{tabular}
	\vspace{2mm}
	\caption{Enclave's Physical source lines of code.}\label{tab:lines_of_code}
\end{table}

In addition, Table~\ref{tab:time_of_flows} summarize the time for some of the main tasks of the system. In our measures, we used a single machine for the agents, clients and Fabric client processes, and a separate machine (on the same local network) that locally emulates the peers and orderer components of the HyperLedger Fabric in a four peers configuration. Therefore, the measurements of the key-transfer protocol and the client-agent protocol do not include any networking overhead, while the ledger requests include almost all of the networking overhead.
Note that the attestation process should only occur once in the lifetime of an \EC agent and therefore its timing is not critical. The ledger time is the most time consuming component of our design as it occurs on every complaint report. Yet for the MeToo use case this is very manageable. For more stressful workload, other strategies would be necessary, see Section~\ref{sec:logging} for an additional discussion.

\begin{table}[ht!]
	\centering
	\begin{tabular}{|c|c|}
		\hline
		Flow & Time\\
		\hline
		Attestation Phase (including communication with IAS) & $3$ sec\\
		\hline
		Key-Transfer Protocol (no networking overhead) & $50$ ms \\
		\hline
		Client-Agent Protocol (no networking \& ledger overhead) & $25$ ms \\
		\hline
		Ledger GET request & $0.3$ sec \\
		\hline
		Ledger UPDATE request & $2.8$ sec \\
		\hline
	\end{tabular}
	\vspace{2mm}
	\caption{Measured time of the main flows}\label{tab:time_of_flows}
\end{table}

\subsection{Some Implementation Challenges}\label{sec:challenges}
Our experience with Intel SGX is that it is not easy to work with. Many of the basic building blocks are either not ready or not applicable to many applications. Other than the inherent limitations such as memory limitations, linkage and porting of simple operations to run inside enclaves are sometime very challenging. Approaches such as \cite{Scone16,Graphene17} make porting easier but at the cost of performance and a significantly larger code base.
We describe two examples of challenges that we encountered in the work.

\mparagraph{Enclave-to-Enclave attestation:}\label{sec:enc-to-enc-attest}
Intel provides end-to-end example code for the remote attestation process \cite{attest-sample-code}.
In their implementation, a client application verifies an enclave and establishes a secure-channel with it using a so called trusted Service-Provider entity. The main problem with this implementation is that it combines both the verification of the enclave and the creation of secure-channel in one protocol.
This means that any new entity that wants to establish a secure channel with the enclave needs to run the entire attestation process with the IAS again, or alternatively put trust in the service provider.
More importantly, it disallows mutual attestation between two enclaves without trust in the service provider.
Instead, we create a simpler attestation code that runs the attestation process with the IAS only once, producing a reusable attestation report. This report can then be used to establish trust and a secure channel with the enclave by any user that is familiar with the IAS public key. A similar approach was recently implemented for forming TLS channels with enclaves \cite{attestTLS18}. For our work we also implemented a version of the attestation report verification that runs inside an enclave, which allows a mutual attestation between enclaves and the creation of a secure channel between two enclaves, which is crucial for passing of the $\econfk$  (See Figure~\ref{fig:transfer_econf_key_flow}).
Our implementation of attestation is open sourced and available at (--blinded for submission--).

\mparagraph{Intel's Trusted-Time Restrictions:}\label{sec:trusted-time}
As mentioned in \cite{trusted-time-and-monotonic}, trusted services such as timer are not supported by the Intel SGX technology itself. Instead, the Intel SGX software provides an implementation of SGX Platform Services that establishes a secure channel with another component, called the Converged Security and Management Engine (CSME), which provides trusted time but is located separately from the CPU.
This is the mechanism that we use for establishing time within enclaves (as part of the time leases mechanism, see Section \ref{sec:cluster-inception}).  Unfortunately, while Intel SGX is supported in Xeon E3 platforms, there is no CSME support there as of today. So until this limitation will be fixed, our current implementation cannot be executed on Xeon E3 hardware.

\section{Related Work}\label{sec:related}
\mparagraph{Enclaves for Secure Computation}
The notion of using SGX enclaves as trusted neutral mediators between parties has been suggested in a number of applications and we mention a partial list here.
Multiparty computation of specific machine learning functions using SGX was studied by Ohrimenko et al.\ \cite{OSF+16ML}. Two-party secure computations where suggested by Gupta et al.\ \cite{GBF+16twoP}. K\"{u}\c{c}\"{u}k et al.\ \cite{KAM+16} used enclaves as a trusted third party for a metering use case. Secure multi-party computations were studied and formalized in the context of SGX by Pass et al. and Bahmani et al.\ \cite{PST17,bahmani2016secure}. Functional encryption using SGX was studied by Fisch et al.\ (Iron)~\cite{Iron17}. Princess~\cite{Princess17} is a collaboration framework for analyzing rare diseases using SGX for privacy preserving collaboration. \cite{BCKS18} hide chaincode operations inside a Blockchain environment. The above studies differ from ours in that they don't consider the enclaves as data owners, but rather as temporarily given access to the data. They also do not consider a cluster of enclaves. Perhaps closest to our work is Iron~\cite{Iron17}, which does consider interaction between enclaves in which one provisions the key to another which computes a function (similar to our core-reader interaction). Their work however focuses on emulating a Functional Encryption setting, where there is a central trusted key authority.

There have also been works studying the use of an enclaves for securing sensitive data in various applications. SecureKeeper~\cite{SecureKeeper16} implements a secured version of Zookeeper in which, like our solution, a number of enclaves store the data using one common secure key. However, unlike our work, this key is provisioned by a central trusted authority.
Brekalo et al.\ and Krawiecka et al.\ \cite{BHS+2016passwd, KPA2016passwd} use enclaves to improve the security of password repositories using techniques that are specific to this use case.
Enclaves have also been used by SGX-Log~\cite{KBLK17Log} to improve the reliability of logging mechanism.

\mparagraph{Rollback prevention in enclaves:} There have been several approaches to preventing state rollback for enclaves. ROTE \cite{ROTE17} uses a cluster of enclaves that communicate to maintain state, while \cite{BCLK17} distribute this task among clients of a service. \cite{BCKS18} addresses rollback issues within a HyperLedger Fabric cluster and builds on trust within this framework. Our approach is close but builds on an administrator lead ledger to ensure both a consistent joint state (that also cannot be rolled back).

\mparagraph{Alternative approaches to the MeToo use case:}
As mentioned in the introduction, Callisto \cite{callisto} is an organization that implemented the two-to-MeToo approach. Their initial deployment relies on full trust in the administrators. Callisto has been working to remedy this and have recently announced a new design \cite{rajan2018callisto} that distributes some of this trust. This design uses a secure key service holding a single key and a cryptographic primitive called OPRF so that clients can register their encrypted complaints without the key server learning it. The complaint is then registered with a DB component. However, compromise of the key server renders their implementation susceptible to enumeration attacks. Compromise of the DB leaks information about whether matches occurred and may be susceptible to erasure of old complaints. Our work presents a stronger security notion We view our work as complimentary, since it provides additional and stronger security guarantees, and for instance, we can use \EC like infrastructure to implement the key server.

\section{Concluding Remarks}
We introduced the idea of autonomous data management by utilizing secure hardware such as Intel SGX and discuss its applicability and merits for specific applications. Our work demonstrates such a design is feasible and we build base tools to do this. We implemented the system for the MeToo use case that achieves much stronger security properties than previous designs and note that each application requires tailoring according to its threat model and security requirements. We believe this notion is a promising building block for mitigating insider attacks in a wide range of applications.

\bibliographystyle{ACM-Reference-Format}
\bibliography{sgx}
\appendix
\section{On the Security of \EC}\label{sec:sec}
in Section~\ref{sec:goals} we describe the general security goals and attributes that our framework achieves.
In the following we describe the security guarantees that our design provides for the MeToo use case (rather than a generic use case). Each guarantee depends on a different set of assumptions and these will made clear.

We view the \EC system as an environment that includes clients, administrators and an authority entity.
The Clients report complaints to the system and have only access to their view of the protocol with the \EC enclave.
Administrators have full access to the network and nodes holding the enclaves. In addition each administrator can view its secluded computation nodes that include their respective secret key and their processes running the Ledger code.
Finally, the authorities receive a periodic encrypted message from an \EC agent running in a scanner role.

Our guarantees rely on the following assumptions:
\begin{itemize}
\item Intel SGX HW is secure. Specifically it provides hiding of all memory of its process and indeed runs the code with a hash according to the attestation report.\footnote{Formal definitions of SGX HW security are complex and we refer the reader to \cite{PST17,Iron17} for examples.}
\item the IAS public key is assumed to be public knowledge.
\item The code compiled into $\econfenclaveso$ indeed implements our design as stated, and the measurement of this library $\econfmrenclave$ is assumed to be public knowledge.
\item The administrator's public keys are assumed to be public knowledge (this assumption is only critical for Property~\ref{pro:match} below).
\end{itemize}

\mparagraph{Complaint secrecy:}This is the most critical property that our design aim to achieve. It basically says that complaints remain secret even if all administrators are corrupted or hacked.
\begin{property}[Complaint secrecy]\label{pro:secrecy}
 An adversary that can corrupt all administrators cannot gain any (non-negligible) advantage in distinguish a true complaint from a random complaint.
\end{property}

\mparagraph{Matching complaints reporting:} This property complements the secrecy property with a guarantee on the actual functionality of the system.

\begin{property}[Non-match Functionality]\label{pro:nonmatch}
 A complaint is never included in the message to the authorities unless there was a matching complaint against the same perpetrator. This holds also in the presence of an adversary that can corrupt all administrators.
\end{property}

\begin{property}[Match Functionality]\label{pro:match}
 If a complaint was accepted by an \EC agent that matches a previous complaint, that was accepted gainst the same perpetrator, then this match will be included in the next message sent to the authorities. This holds as long as at least $k$ out of $n$ administrators are not corrupted.
\end{property}

\section{Additional Implementation Details}\label{app:details}

\subsection{Ensuring Key and Data Survivability}\label{sec:survival}
An important attribute of our system is its ability to survive hardware failures, and most crucially, not to lose the complaints data. Moreover, we want to enable the survivability property in face of a denial of service attack, as long as enough administrators are not corrupted.
There are two aspects to this: First there is the encrypted data in the ledger. This is straightforward to handle within the HyperLedger Fabric framework. Since each administrator holds a copy of the data, and can also replicate it by adding more peers (possibly in remote locations), we can ensure the data will be retrievable by at least $k$ of the $n$ administrators even in face of failures.

The second aspect is the key survivability, since the encrypted data is useless without an enclave holding the $\econfk$. To this end, we have multiple \EC agents in the system, some will remain active even in face of hardware failures. In order to ensure survival in face of an internal attack that erases all of the agents (e.g. by a rogue administrator), we advocate that each administrator hold a back up \EC agent in his personal domain. This backup agent can receive leases just for the initial phase in which it obtains the $\econfk$, and then its leases only need to be renewed in case of loss of the other live agents.
Note that leases can be renewed after being stale. One subtlety is that the trusted time in an enclave relies on a nonce which is retrieved during the attestation phase. Yet, for an existing enclave, we added an API for retrieving the trusted time from an agent whose leases are already stale. This allows the revival of leases for agents whose hardware has been upgraded and the time nonce has been changed.

\subsection{Upgrading Agents}\label{sec:upgrade}
The ability to upgrade the enclave code is tricky. On the one hand it undermines the entire security premise of the system -- that no human entity be able to retrieve the $\econfk$. This property builds strongly on the fact that agent enclaves only hand out the $\econfk$ to other enclaves that run the exact same code.
However, in reality it is very likely that bug fixing or feature addition would be required (even in the MeToo use case where the functionality is very simple). To handle upgrades, one needs to build into the initial enclave code an override in which if a sufficient number of {\em special leases} are given to an enclave, then it can receive the $\econfk$ even with a different $\mrenclave$. Such a scenario  should be very rare and should require leases from more than the regular administrator, e.g., have additional consent from high ranking managers in the organization. The criteria for such a special upgrade should be hardcoded in the enclave code.

\subsection{Logging Mechanism}\label{sec:logging}

We implemented a trusted logging mechanism that tracks the enclave's operations. More precisely, in each ECALL the agent-enclave also returns a log record that contains a (trusted) time, message and a signature over it's new state, where the state is defined by an aggregated hash over all previous log records. We consider three-types of log messages: 1) An INFO message that notifies about an event that occurred (\eg the reception of a complaint). 2) A WARNING/ERROR message that notifies about an unexpected behavior, and 3) A PERIODIC message that the untrusted part periodically retrieves from the enclave (\eg once an hour) using a special ECALL. The untrusted part is responsible for writing these logs into a local storage, where these logs are mainly used by the administrator's software for the decision of whether to renew the lease or not. For example, in our implementation, before renewing a lease of an agent, the administrator's software first verifies all signatures, then it checks whether there are no ERROR messages and finally it verifies that indeed there is a verified PERIODIC message in every hour with respect the agent's trusted time. Finally, it renews the lease with respect to the trusted time that appears in the last log record.

The logging mechanism described above has the limitation that logs might be erased by an adversarial administrator. Such an erasure can be detected by the other administrators by monitoring that there are no missing PERIODIC logs. In addition the administrators need to ensure that the existing logs do not include indication of an unexpected restart of the enclave, which is an indication of possible replay attack on the enclave. This serves as a mechanism to detect foul play and not renew a lease accordingly. However, there is no mechanism to retrieve the content of erased logs.
For critical information, we therefore advise to register it in the immutable ledger, for which data is not lost as long as enough administrators remain loyal. However, due to the performance limitations of the ledger, the amount of log data added to the ledger should be limited and contain only critical information. One example is to log every restart of an enclave to the ledger so that multiple restarts can be avoided.


\subsection{Scanner Agent}\label{sec:scanner-agent}

As mentioned in Section~\ref{sec:arch}, the scanner agent is responsible for performing a periodic scan on the encrypted complaints in the ledger for finding matches. We implemented a naive scanner that maintains inside the enclave's memory a map table $M$ that maps a perpetrator-id to the complaints it appears in, together with a list $R$ that contains all the perpetrator-id's that has (potential) new match. On every scan, the agent requests for the entire content of the ledger using a GET request. Then, using the $\econfk$, the enclave starts to decrypt each encrypted complaint from the point it stopped the previous scan. On every decrypted complaint, it extracts the perpetrator-id and the user-id and updates the corresponding entry in $M$ (\ie updates the relevant entry in $M$ in case there is no previous complaint on perpetrator-id from the same user-id). In case of a match, the enclave inserts perpetrator-id into $R$. At the end of the scan, the enclave prepares a report including all matches in $R$ and encrypts it for the authorities (using an additional hard-coded authorities public-key). The report is always prepared and is of a fixed size, even if $R$ is empty (\ie no new match was found). This makes report that contains matches indistinguishable from an empty report. Once the report is sent, the enclave prepares a special encrypted record and the agent inserts it to the ledger using the UPDATE request. This special record includes an indicator of the last complaint that was scanned for matches so that a future scan would not repeat old matches. Observe that in case of a large amount of complaints, the map $M$ might be large and might exceed the size of the enclave's page cache (128MB). In such case, the scanning will be more time consuming due to paging overhead. Since this process is done offline (and once a day), this potential overhead is not significant to our system.

\subsection{About the IAS Report}\label{sec:about-ias-report}
The IAS's public-key $\iaspk$ is of type RSA3072 and appears as part of their attestation CA-certificate (PEM format). This certificate is published at Intel's website and is sent to any new user as part of the registration process. In our implementation, the IAS-report consists of three fields that are actually contained in any response from IAS. The fields are: report\_body ($\approx1$KB), cert\_chain ($\approx3.5$KB) and signature ($\approx350$ bytes). The report\_body is in JSON format. It has many variables, including the variable ``isvEnclaveQuoteBody'' which contains the quote in (URL-safe) base64 format. The cert\_chain can be verified using Intel's public CA-certificate, and it contains an RSA2048 public-key. Finally, the signature that has been taken over the report\_body, can be verified using the RSA2048 public key. See \cite{IAS-API} for more details about IAS.

\end{document}